\documentclass[useAMS,usenatbib]{mn2e}
\usepackage{epsfig}
\usepackage{float}

\usepackage{natbib}

\usepackage{color}

\usepackage{aas_macros}

\usepackage{array}
\newcolumntype{L}[1]{>{\raggedright\let\newline\\\arraybackslash\hspace{0pt}}m{#1}}
\newcolumntype{C}[1]{>{\centering\let\newline\\\arraybackslash\hspace{0pt}}m{#1}}
\newcolumntype{R}[1]{>{\raggedleft\let\newline\\\arraybackslash\hspace{0pt}}m{#1}}

\def\sigs{\mbox{$\sigma_\star$}}

\def\Re{\mbox{$R_{\rm e}$}}

\def\Msun{\mbox{$M_\odot$}}
\def\Mtot{\mbox{$M_{\rm tot}$}}

\def\ML{\mbox{$M/L$}}
\def\dimf{\mbox{$\delta_{\rm IMF}$}}
\def\Yst{\mbox{$\Upsilon_\star$}}

\def\Mdyn{\mbox{$M_{\rm dyn}$}}

\def\mst{\mbox{$M_{\star}$}}

\def\Mvir{\mbox{$M_{\rm vir}$}}
\def\cvir{\mbox{$c_{\rm vir}$}}

\def\fdm{\mbox{$f_{\rm DM}$}}

\def\lsim{\mathrel{\rlap{\lower3.5pt\hbox{\hskip0.5pt$\sim$}}
    \raise0.5pt\hbox{$<$}}}                
\def\gsim{~\rlap{$>$}{\lower 1.0ex\hbox{$\sim$}}}
\def\Rap{\mbox{$R_{\rm Ap}$}}

\def\sigAp{\mbox{$\sigma_{\rm Ap}$}}
\def\sige{\mbox{$\sigma_{\rm e}$}}

\def\atlas3d{ATLAS$^{\rm 3D}$}

\def\al{\mbox{$\alpha_{\rm l}$}}
\def\alRe{$\alpha_{\rm l}(\Re)$}
\def\alRetwo{\mbox{$\alpha_{\rm l}(\Re/2)$}}

\def\amw{\mbox{$\alpha_{\rm mw}$}}
\def\amwRe{\mbox{$\alpha_{\rm mw}(\Re)$}}

\title[Mass density slopes in ETGs]{Systematic variations of central mass density slopes in early-type galaxies}

\author[Tortora C. et al.]{\noindent
C.~Tortora$^{1}$\thanks{E-mail: ctortora@na.astro.it}, F.~La
Barbera$^{1}$,  N.R. Napolitano$^{1}$, A.J.~Romanowsky$^{2,3}$,
I.~Ferreras$^{4}$, \newauthor R.~R. de Carvalho$^{5}$
\\~\\
$^{1}$ INAF -- Osservatorio Astronomico di Capodimonte, Salita
Moiariello, 16, 80131 - Napoli, Italy \\
$^2$ Department of Physics and Astronomy, San Jos\'e State
University, San Jose, CA 95192, USA\\
$^3$ University of California Observatories, 1156 High Street,
Santa Cruz, CA 95064, USA\\
$^4$ Mullard Space Science Laboratory, University College London,
Holmbury St Mary, Dorking, Surrey RH5 6NT\\
$^5$ Instituto Nacional de Pesquisas Espaciais / MCTI Av. dos
Astronautas 1758, Jd. Granja S$\rm \tilde{ã}$o Jos\'e dos Campos -
12227-010 SP}

\begin{document}
\date{Accepted  Received }
\pagerange{\pageref{firstpage}--\pageref{lastpage}} \pubyear{xxxx}
\maketitle

\label{firstpage}
\begin{abstract}
We study the total density distribution in the central regions
($\lsim \, 1$ effective radius, \Re) of early-type galaxies
(ETGs), using data from SPIDER and \atlas3d. Our analysis extends
the range of galaxy stellar mass ($\mst$) probed by gravitational
lensing, down to $\sim 10^{10}\, \rm \Msun$. We model each galaxy
with two components (dark matter halo + stars), exploring
different assumptions for the dark matter (DM) halo profile (i.e.
NFW, NFW-contracted, and Burkert profiles), and leaving stellar
mass-to-light (\mst$/L$) ratios as free fitting parameters to the
data. For all plausible halo models, the best-fitting \mst$/L$,
normalized to that for a Chabrier IMF, increases systematically
with galaxy size and mass. For an NFW profile, the slope of the
total mass profile is non-universal, independently of several
ingredients in the modeling (e.g., halo contraction, anisotropy,
and rotation velocity in ETGs). For the most massive ($\mst \sim
10^{11.5} \, M_\odot$) or largest ($\Re \sim 15 \, \rm kpc$) ETGs,
the profile is isothermal in the central regions ($\sim \Re/2$),
while for the low-mass ($\mst \sim 10^{10.2} \, M_\odot$) or
smallest ($\Re \sim 0.5 \, \rm kpc$) systems, the profile is
steeper than isothermal, with slopes similar to those for a
constant-\ML\ profile. For a steeper concentration-mass relation
than that expected from simulations, the correlation of density
slope with galaxy mass tends to flatten, while correlations with
\Re\ and velocity dispersions are more robust. Our results clearly
point to a ``non-homology'' in the total mass distribution of
ETGs, which simulations of galaxy formation suggest may be related
to a varying role of dissipation with galaxy mass.
\end{abstract}


\section{Introduction}\label{sec:intro}

In the so-called $\Lambda$CDM model, the formation of virialized
dark matter (DM) haloes proceeds from the initial expansion and
subsequent collapse of tiny density perturbations. The assembly
and evolution of DM haloes can be studied in detail by means of
N-body simulations, which predict that the DM density profile,
$\rho_{\rm DM}(r)$, should be independent of halo mass, being well
described by a double power-law relation -- the so-called NFW
profile -- with $\rho_{\rm DM}(r) \propto r^{-3}$ in the outer
regions, and $\rho_{\rm DM}(r) \propto r^{\alpha}$, with $\alpha <
0$, in the centre ($\alpha=-1$, \citealt{NFW96}; $\alpha=-1.5$
\citealt{Moore+98}). However, N-body simulations follow only the
evolution of DM particles, not including  the extremely complex
physics of gas and stars. These components are dominant in the
central regions of galaxies, in particular those of early-type
galaxies  (ETGs), which exhibit a peaked surface brightness
profile, typically well described by the S\'ersic law
(\citealt{Ciotti91}), with a shape parameter, $n$ (S\'ersic
index), that accounts for variations of the light profile shape
among galaxies. Gas and stars, falling down into the DM potential
well, could drag a significant amount of DM particles inside,
making the DM profile more concentrated in the galaxy centre
(\citealt{Blumenthal+86}; \citealt{Gnedin+04};
\citealt{DelPopolo09}; \citealt{Cardone+11SIM}) than the
``expected'' NFW law. The study of the DM profile in the inner
regions of ETGs is also hampered by the degeneracy between the
shape of the DM profile and that of the stellar Initial Mass
Function (IMF; \citealt{NRT10}; \citealt{Dutton+11a, Dutton+12a};
\citealt{TRN13_SPIDER_IMF}).

For massive galaxies ($\mst \sim 10^{11-12} \, M_\odot$),
gravitational lensing and studies of stellar dynamics in the
galaxy central regions have found that the light and halo profiles
conspire to have a total mass density profile which is nearly
isothermal (\citealt{Bolton+06_SLACSI};
\citealt{Bolton+08_SLACSV}; \citealt{Auger+09_SLACSIX};
\citealt{Auger+10_SLACSX}; \citealt{Chae+14}; \citealt{Oguri+14}),
i.e. a total-mass density slope of $\alpha \sim -2$. The crucial
questions here are why there is such a conspiracy, and if it is
universal (i.e. holds for all galactic systems). Indeed,
observations suggest that this is not the case. { For low-mass
ETGs, \cite{Dutton_Treu14} have recently shown that density
profiles are steeper than isothermal, while they are isothermal
for high-mass systems. At the mass scale of groups and clusters of
galaxies, the mass density distributions appear to be also
shallower than isothermal (e.g. \citealt{Sand+08};
\citealt{Humphrey_Buote10}). \citet{deBlok+01} also found
shallower-than-isothermal profiles in low surface-brightness
galaxies (with $\alpha \sim 0$). The recent theoretical work by
\cite{Remus+13} seems to provide a theoretical framework to
interpret these results. Their simulations show that {\it in-situ}
star formation, resulting from dissipative processes, tends to
form steeper-than-isothermal profiles, while gas-poor mergers are
a natural attractor towards the isothermal slope. This motivates
for further, in-depth, studies of the slope of total-mass density
profiles and their correlations with galaxy properties, while
making connections  to the theory.

In the present work, we study the slope of the mass density
profile of ETGs in a wide mass range, using data from the
SDSS-based SPIDER survey (\citealt{SPIDER-I}), one of the largest
well-characterized samples of ETGs in the nearby Universe -- with
high-quality spectroscopy and optical plus Near-Infrared (NIR)
photometry available -- as well as the \atlas3d\ sample
(\citealt{Cappellari+11_ATLAS3D_I}). We probe the galaxy mass
profiles down to a stellar mass of $10^{10} \, M_\odot$, hence
extending, with an independent approach, results of gravitational
lensing studies for massive galaxies (\citealt{Bolton+06_SLACSI};
\citealt{Bolton+08_SLACSV}; \citealt{Auger+09_SLACSIX};
\citealt{Auger+10_SLACSX}). We perform a Jeans dynamical analysis
of the available photometric and spectroscopic data, with a suite
of dynamical models (see \citealt{TRN13_SPIDER_IMF}), testing
several assumptions on the shape of the DM halo profile and
leaving stellar mass-to-light ratios as free parameters in the
analysis. Our work complements previous studies (e.g.
\citealt{Humphrey_Buote10}; \citealt{Dutton_Treu14}), in that it
compares findings for two independent, well-characterized, samples
of ETGs, and investigates the impact of a variety of modeling
ingredients. Our goal is to scrutinize if the central density
slope of the total mass distribution in ETGs stays isothermal or
changes with mass and other galaxy properties, comparing to
predictions of simulations of galaxy formation.

The paper is organized as follows. In Sec.~\ref{sec:data} we
present the samples of ETGs used in the present study.
Sec.~\ref{sec:analysis} deals with data analysis, describing the
dynamical modeling approach and our definitions of the mass
density slope. Sec.~\ref{sec:IMF_trends} and~\ref{sec:results}
present the correlations of the mass-to-light ratio and total mass
density slope with galaxy properties, i.e. structural parameters,
velocity dispersion, stellar mass, and DM fraction. In
Sec.~\ref{sec:literature}, we present an extensive comparison of
our results with those from the literature.
Sec.~\ref{sec:conclusions} summarizes results and conclusions.

\section{Samples}\label{sec:data}

We rely on two samples of ETGs, one main sample
from the SPIDER survey (\citealt{SPIDER-I}), and  a complementary data-set
from the \atlas3d\ project (\citealt{Cappellari+11_ATLAS3D_I}).

\subsection{SPIDER sample}
\label{sec:sample_SPIDER} The SPIDER data-set is described in
\citet{SPIDER-I}. It consists of a sample of $5,080$ bright
($M_r<-20$) ETGs, in the redshift range of  $z=0.05$ to $0.095$,
with optical and NIR photometry available ($grizYJHK$ wavebands)
from the Sloan Digital Sky Survey (SDSS) DR6 and the UKIRT
Infrared Deep Sky Survey-Large Area Survey
DR3~\footnote{http://www.sdss.org, http://www.ukidss.org}.
Structural parameters, i.e. the effective radius \Re\ and S\'ersic
index $n$, have been measured homogeneously for all galaxies, from
$g$ through $K$, using the software 2DPHOT
(\citealt{LaBarbera_08_2DPHOT}). SPIDER ETGs have central velocity
dispersions, \sigAp, available from the SDSS, measured in the
circular aperture of the SDSS fiber (having radius $\Rap = 1.5$ arcsec). {
The median ratio of the SDSS fibre to the K-band
effective radius, $\Rap/\Re$, amounts to $\sim 0.6$, implying only
a mild extrapolation in the estimate of mass density slopes (see below).}

ETGs are defined  as  bulge-dominated systems  (i.e. SDSS
parameter $fracDev_r \!  > \!  0.8$, where $fracDev_r$ measures
the fraction of galaxy light better fitted by a de~Vaucouleurs,
rather than an exponential law), featuring passive spectra within
the SDSS fibres (SDSS attribute $eClass \!   < \!  0$, where
$eClass$ indicates the  spectral  type  of  a  galaxy  based on  a
principal  component analysis). For the present work, we rely on a
subsample of $\sim 4300$ SPIDER ETGs, with better quality optical
and NIR structural parameters, selected as in \citet{SPIDER-VI}.
For each galaxy, the stellar mass-to-light ratio, \Yst, has been
estimated by fitting \citet{BC03} stellar population models to the
multi-band photometry, under the assumption of a Chabrier IMF (see
\cite{SPIDER-V} and \cite{SPIDER-VI} for details). The sample is
$95\%$ complete down to a stellar mass of $\mst \sim 3 \times
10^{10}\, \rm \Msun$.

\subsection{\atlas3d\ sample}
\label{sec:sample_atlas3d}

Our second sample consists of 260  ETGs from the \atlas3d\
survey (\citealt{Cappellari+13_ATLAS3D_XV,
Cappellari+13_ATLAS3D_XX}). Further details about the  selection
of  \atlas3d\ galaxies are provided in \cite{Tortora+14_MOND}. For
each galaxy, we perform the dynamical analysis by using (i) its
$r$-band effective radius, \Re, (ii) the $r$-band total luminosity
$L_r$, (iii) the projected stellar velocity dispersion, \sige,
within \Re, and (iv) the stellar mass-to-light ratio
($\Upsilon_*$) derived by fitting galaxy spectra with
\cite{Vazdekis+12} single SSP MILES models, having a
\cite{Salpeter55} IMF. Stellar masses are converted to a
\cite{Chabrier01} IMF, using the fact that the Chabrier IMF
normalization is $\sim 0.26$ dex smaller than the Salpeter one.
We notice that out of 260 ETGs, about 15\% of \atlas3d\ ETGs have significant stellar
mass-to-light ratio gradients and young stellar populations (with
an H$\beta$ equivalent width $>$2.3~\AA). We found that excluding  these
objects from the analysis does not affect at all the trends of total-mass density slope.

As discussed in \cite{Tortora+14_MOND}, it is important to note
that the published $L_r$ and \Re\ values are not self-consistent.
The former correspond to detailed multi-gaussian expansion (MGE)
fits that were truncated at typically $\sim$~4~\Re. The latter are
the MGE-based values renormalized by a factor of 1.35 to
correspond to more conventional estimates from the literature.
Here we will use these \Re\ values, but adjust each $L_r$ value
such that the projected luminosity inside \Re\ for our adopted de
Vaucouleurs model is the same as in the original MGE model.  This
means increasing $L_r$ by typically a factor of 1.2.

\section{Analysis}\label{sec:analysis}

\subsection{Dynamical modeling}

We derive the dynamical (i.e. total) mass distribution of ETGs by
solving spherical isotropic Jeans equations, where a given model
for the mass profile is fitted to \sigAp\ and \sige, for SPIDER
and \atlas3d\ ETGs, respectively. We use two-component mass
models, describing stars and DM.

The stellar mass profile is modeled by either a S\'ersic (SPIDER)
or a \cite{deVauc48} (\atlas3d) law. The shape parameter $n$ and
effective radius of the S\'ersic laws are those obtained by
fitting galaxy images in K band (see Sec.~\ref{sec:data}). For the
de Vaucouleurs law, we use r-band effective radii from the
\atlas3d\ sample. In both cases, the total luminosity of the light
distribution is converted into stellar mass  by means of the
stellar mass-to-light ratio,  \Yst, which is a free fitting
parameter \footnote{{Note that stellar mass-to-light ratios
estimated from stellar population models (Sec.~\ref{sec:data}) are
not used to derive the density slopes, but only as reference
values to normalize the best-fitting \Yst s, and to produce
correlations of density slope with stellar-mass estimates for a
``standard'' (i.e. MW-like) IMF (allowing a more direct comparison
to other studies).}}. This procedure assumes that the shape and
scale radius of the stellar mass distribution of ETGs are the same
as for the light distribution, i.e. one can neglect \Yst\
gradients inside these galaxies.  {One can notice that although
S\'ersic fits are known to provide a better fit to the light
distribution of ETGs than a pure de Vaucouleurs law, a comparison
of results for S\'ersic (SPIDER) vs. de Vaucouleurs (\atlas3d)
profiles is useful to test the robustness of the results against
the parametrization of the galaxy light distribution. Also, K-band
light is more sensitive to the old quiescent component of an
unresolved stellar population, describing more closely the stellar
mass profile of a galaxy, than the light distribution at optical
wavebands. Hence, the comparison of K- (SPIDER) and r- (\atlas3d)
band results allows us to test the impact of stellar mass-to-light
ratio gradients in galaxies, besides that of selecting two
different samples of ETGs, and using different parametrizations of
the galaxy light profiles. As a further test, we also compare our
findings to those obtained for the (same) SPIDER sample using
(SDSS-based) de Vaucouleurs (rather than S\'ersic) structural
parameters in the r (rather than K) band. In general, as discussed
below, assuming a constant \Yst\ does not likely affect
significantly our conclusions.

For the DM component, in the case of the SDSS-based SPIDER sample
we can rely only on velocity  dispersions measured within a single
aperture (i.e. the SDSS fibre), which does not allow us to
constrain the shape of the DM profile in detail. In contrast,
using the spatially extended kinematics of \atlas3d\ galaxies, one
could constrain, in principle, { the shape of both the stellar and
DM components in the central galaxy regions} in detail, as shown,
e.g., in~\cite{Cappellari+12}. In the present work, to perform a
clean comparison of results from both samples, we apply the same
procedure to both SPIDER and \atlas3d\ ETGs, fitting two-component
models to central velocity dispersion estimates for both samples.
To this effect, we explore a variety of models for the DM
component, exploring several plausible assumptions
(\citealt{TRN13_SPIDER_IMF}).

\begin{figure*}
\psfig{file=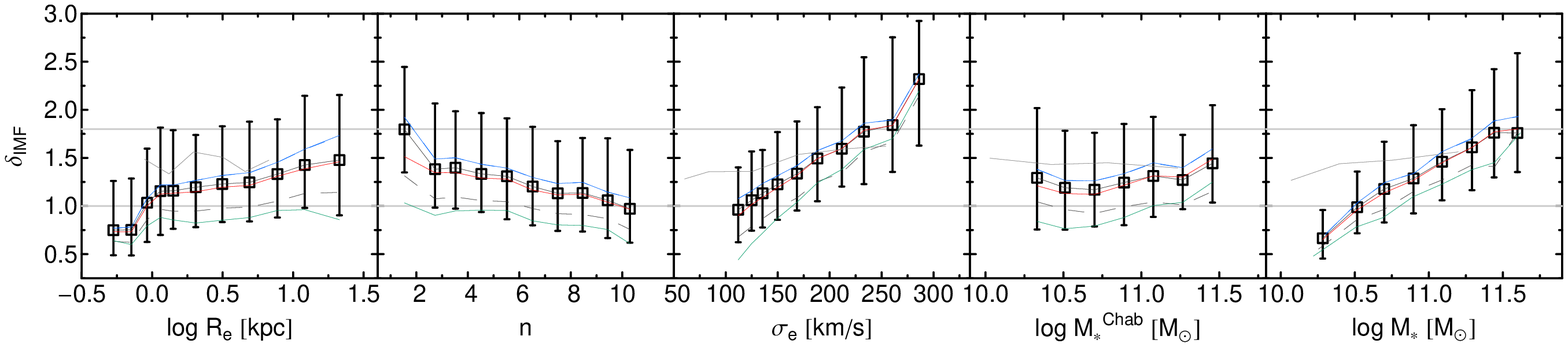, width=1.01\textwidth} \caption{IMF mismatch
parameter, \dimf, for our sample of SPIDER ETGs,  as a function of
(from left to right) \Re, $n$, \sige, $\mst^{\rm Chab}$ (estimated
for a Chabrier IMF) and \mst\ (allowing for a variable IMF
normalization). Open squares and error bars are median and
16--84th percentile trends for our fiducial NFW+S\'ersic galaxy
model. We also plot results for NFW models with fixed $\Mvir =
10^{13} \, \rm \Mvir$ (red curve), ``high-concentration'' NFW
models (green), contracted NFW models (dashed black line), and
Burkert profiles with $r_{\rm B} = 1 \, \rm kpc$ (solid blue
line). Notice that results for Burkert models with $r_{\rm B} = 20
\, \rm kpc$ are not shown in the plot, as the corresponding trends
are identical to the case for $r_{\rm B} = 1 \, \rm kpc$. The
results for \atlas3d\ using the fiducial NFW+S\'ersic galaxy model
are plotted as dark gray lines. Light gray horizontal lines mark
the \dimf\ values expected for a Chabrier (\dimf$=1$) and Salpeter
(\dimf$\sim1.8$) IMF.}\label{fig:dIMF}
\end{figure*}

\begin{description}
\item \emph{\citet[hereafter NFW]{NFW96} profiles}. {The DM profile from
N-body simulations is well described by a double power-law,
commonly referred to as the NFW profile (\citealt{NFW96, NFW97}).
In the present work, we adopt the NFW as the reference DM profile,
assuming also the correlation between virial mass and
concentration (\Mvir\ and \cvir , respectively), that applies to a
WMAP5 cosmology \citep{Maccio+08} as well as the \Mvir--$\mst^{\rm
Chab}$ correlation from \citet{Moster+10}. In order to explore the
effect of a possible modification to the DM profile because of the
interaction between gas and stars with DM, we also consider the
case of an NFW with an adjustable degree of baryon-induced
adiabatic contraction (AC, \citealt{Gnedin+04}). Also, we explore
how our results depend on the assumed \Mvir -- \cvir\ relation, by
(i) adopting a constant \Mvir $= 10^{13}\, \rm \Msun$ (and
deriving the corresponding, constant, \cvir\ from the
\citealt{Maccio+08} \Mvir --\cvir\ relation), and (ii) using the
\cvir -- \Mvir\ correlation, based on observations, from
\citet[hereafter LFS12]{Leier+12}. In the latter case, we adopt
the relation obtained from LFS12  by combining X-ray results
from~\citet{Buote+07} with a gravitational lensing analysis of
galaxies at intermediate redshifts~\footnote{We used the Eq.~(11)
from LFS12, with the  {\it comb} best-fitting slope and
normalization coefficients from their Table~1.}. For the median
redshift of the SPIDER sample ($z \sim 0.08$), the relation is
written as $\cvir = 9.62 \times \Mvir^{-0.278}$. Notice that
although LFS12 found some evidence for a variation of the slope of
the \Mvir -- \cvir\ relation with the \Mvir\ range fitted, this is
unimportant for the relatively narrow mass range covered by our
sample of ETGs with respect to that of LFS12. As discussed below,
the LFS12 relation is significantly steeper than the
\citet{Maccio+08} one, providing significantly higher
concentrations for the lowest mass galaxies analyzed in this work.
Also, LFS12 assumed a MW-like, Chabrier, IMF to map the stellar
mass distribution of lensing galaxies, while in the present study
we keep the \Yst\ (i.e. the ``IMF normalization'') as a free
fitting parameter. In the following, we refer to models with NFW
profiles and a \cvir -- \Mvir\ relation from LFS12 as
``high-concentration'' NFW models.
 }

\item \emph{\citet{Burkert95} profiles.}
The Burkert profile is the prototype of cored models, and has been
shown to reproduce quite well the DM profile of spirals and dwarf
galaxies. The density and scale parameter of the Burkert profile
($\rho_{\rm B}$ and $r_{\rm B}$, respectively) are assumed to
follow the $\rho_{\rm B}-r_{\rm B}$ relation from
\cite{Salucci_Burkert00}, { adjusted to match results at higher
surface density}, for two ETGs, by \citet[hereafter
MSB11]{Memola+11}. We explore two cases in detail, where the scale
radius is set to $r_{\rm B} = 1$ and $20 \, \rm kpc$,
respectively. The possible impact of a varying $r_{\rm B}$ (with,
e.g., galaxy mass) on our results is discussed in
Sec.~\ref{sec:comp_models}.
\end{description}
For each galaxy and a given DM model, one has one single fitting
parameter, i.e. the mass-to-light ratio \Yst . The \Yst\ is
constrained by solving the Jeans equations to match the available
velocity dispersion estimate (see above). { We have performed
several tests, showing that  our results
are quite independent of the assumptions on the DM profile.
 None of
the conclusions is changed when comparing results for NFW profiles
with either a constant \Mvir $= 10^{13}\, \rm \Msun$ (and constant
\cvir), or the \citet{Maccio+08} \cvir -- \Mvir\ relation.
However, assuming a cored Burkert profile or high concentration
haloes -- consistent with LFS12 -- can affect significantly some
of our results, as discussed below.}

\subsection{Inferring the slope of the density profile}
\label{sec:def_slopes}

We aim here to study the slope of the total  mass profile of ETGs,
rather than that of DM only (as in our previous work,
see~\citealt{NRT10}). For each galaxy, at a given (deprojected)
galacto-centric distance, $r$, the total mass density, $\rho(r)$,
is obtained by summing the best-fitting stellar mass profile and
the DM profile at that radius. In order to probe the robustness of
the correlations between mass-density slope and galaxy properties,
we adopt different definitions of the slope.
\begin{description}
 \item[i) ] We define the local logarithmic slope of the profile, $\al(r)
= d\log \rho(r)/d \log r$;
 \item[ii) ]  We compute the {\it mass-weighted logarithmic
slope}, \amw, within a given radius $r$ (\citealt{Koopmans+09};
\citealt{Dutton_Treu14}). It is defined as:
\begin{equation}
\amw(r) \equiv \frac{1}{M(r)} \int_{0}^{r} \al(r) 4 \pi x^{2}
\rho(x) dx = -3 + \frac{4 \pi r^{3} \rho(r)}{M(r)},
\end{equation}
where $M(r)$ is the (total) mass enclosed within a sphere of
radius $r$.  Some algebra shows that
\begin{equation}
\amw(r) = - 3 + d\log M(r)/d\log r.
\end{equation}
\end{description}

For a power-law density profile, $\rho \propto r^{\alpha}$, one
has $\al(r)=\amw(r)=\alpha$ at all radii, while this is not true
for a general density distribution. However, in general, one can
demonstrate that $\amw(r) > \alpha$. We calculate the logarithmic
slope at two different radii, i.e. $\Re/2$ and $\Re$,
respectively, while the mass-weighted slope is computed within
$r=$\Re . These choices are motivated by the fact that the
available estimates of velocity dispersion refer to an aperture
with radius ranging from a few tenths of \Re\ to about one \Re , for
both the SPIDER and \atlas3d\ samples. Thus, our choice minimizes
the amount of extrapolation of the best-fitting mass models,
exploring at the same time the inner profile at different radii ($\Re/2$ and $\Re$).

In the following sections we will discuss these slopes in terms of
\Re, S\'ersic $n$, \sige , stellar mass and central DM fraction
within a radius $R$, defined as $\fdm (R) = 1 - \mst(R) /
\Mtot(R)$.

\section{Mass-to-light ratio trends}\label{sec:IMF_trends}

Using the dynamical estimate of \Yst\ (see
Sec.~\ref{sec:analysis}), we define the mismatch parameter, $\dimf
= \Yst / \Yst^{\rm Chab}$, where $\Yst^{\rm Chab}$ is the stellar
mass-to-light ratio obtained by fitting data (either colors or
galaxy spectra) with stellar population models having a Chabrier
IMF. The \dimf\ can be interpreted as a variation in the
normalization of the IMF with respect to the case of a
``standard'', Milky-Way-like, distribution.

Fig.~\ref{fig:dIMF} plots the mismatch parameter as a function of
\Re, S\'ersic index $n$, \sige\ and \mst, the latter estimated
with either $\Yst^{\rm Chab}$ (i.e. a Chabrier IMF; \mst$^{\rm
Chab}$) or \Yst\ (i.e., the best-fitting IMF normalization; \mst).
We find that \dimf\ is positively correlated with \Re, \sige , and
\mst, becoming larger than 1 in more massive and bigger galaxies.
In contrast, the \dimf\ decreases with $n$, while it is almost
constant with \mst$^{\rm Chab}$. The trends for Burkert and
``high-concentration'' NFW models encompass the range of values
for all trends. Although the absolute value of \dimf\ depends on
the adopted DM profile in the modeling, the relative trends of
\dimf\ trends with galaxy parameters are robust, being independent
of the assumptions on the DM model (e.g., NFW vs. AC+NFW vs.
Burkert profiles), and the assumed \cvir -- \Mvir\ relation.

The mismatch parameter for \atlas3d\ using the fiducial
NFW+S\'ersic galaxy model is also shown. All the correlations are
shallower, and \Yst\ have values $\sim 15\%$ larger than
SPIDER-based results (e.g. \citealt{TRN13_SPIDER_IMF}).

As noted by~\citet{Cappellari+12}, a \dimf\ larger than one can be
due to either a bottom-heavy IMF (because of the large fraction of
dwarf relative to giant stars) or a top-heavy distribution
(because of the large fraction of stellar remnants from evolved
massive stars). The degeneracy can be broken by studying
gravity-sensitive features in the integrated light of ETGs
\citep{Conroy_vanDokkum12b}. These features allow one to constrain
the mass fraction of dwarf-to-giant stars in the IMF, rather than
the IMF normalization itself~\citep{LaBarbera+13_SPIDERVIII_IMF},
with several studies having found evidence for a  bottom-heavier
than Chabrier IMF, in high- relative to low-$\sigma$
ETGs~\citep{Ferreras+13, Spiniello+14}. As shown in
Fig.~\ref{fig:dIMF}, at large \Re\ and \sige , our results are
consistent with the IMF normalization expected for a Salpeter IMF,
or even a bottom-heavier than Salpeter IMF.

\begin{figure*}
\psfig{file=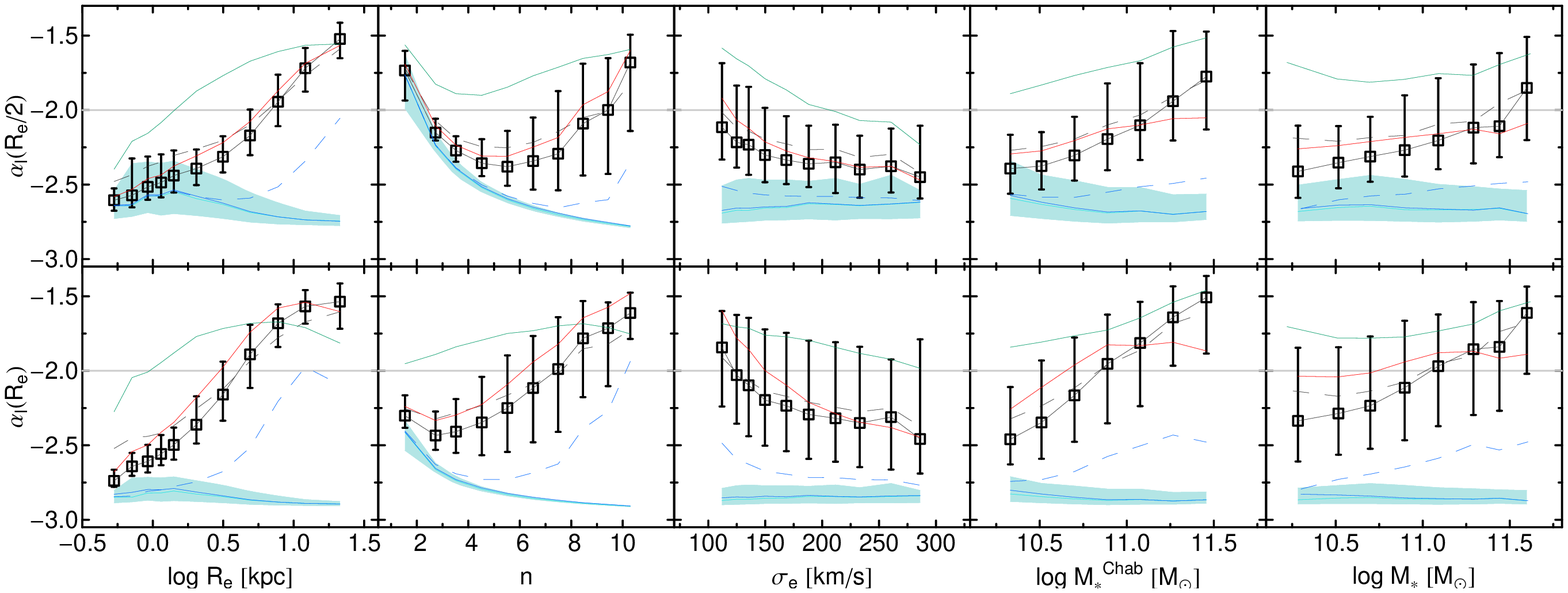, width=1.01\textwidth} \caption{Mass density
slopes, for our sample of SPIDER ETGs,  as a function of (from
left to right) \Re, $n$, \sige, $\mst^{Chab}$ (Chabrier-IMF
stellar mass) and \mst\ (stellar mass estimated allowing for a
variable IMF normalization). Top and bottom panels refer to the
logarithmic density slopes, \alRetwo\ and \alRe , respectively.
All slope values are for models with variable \Yst . Symbols are
as in Fig.~\ref{fig:dIMF}. Results for Burkert profile with
$r_{\rm B} = 1$ and $20\,$ kpc are plotted as  solid blue and
dashed blue lines, respectively. The cyan line and shaded regions
mark median and 16--84th percentile slopes for the stellar mass
distribution only. In all panels, the gray horizontal line marks
the slope value of $-2$, corresponding to the case of an
isothermal sphere.}\label{fig:slopes_1}
\end{figure*}

\begin{figure*}
\psfig{file=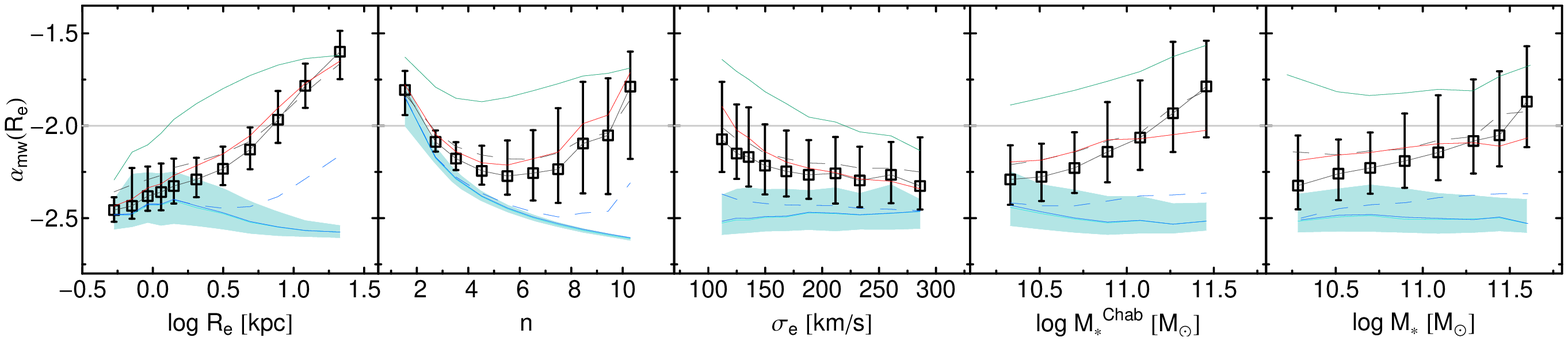, width=1.01\textwidth} \caption{The same as
Fig.~\ref{fig:slopes_1}, but for the mass-weighted density slope
$\amwRe$ .}\label{fig:slopes_2}
\end{figure*}

\begin{table*}
\centering \caption{Best fit parameters with $1\, \sigma$ errors
for the relation \alRetwo\  $= a + b x + c x^{2}$, where $x = \rm
\log \Re/ \rm 3 kpc$, $n/4$, $\sige/ \rm 200 km/s$, $\rm \log
\mst^{\rm Chab}/(10^{11}\Msun)$ and $\rm \log
\mst/(10^{11}\Msun)$. The scatter (rms) around the best fit
relation is also reported.}\label{tab:tab1}
\begin{tabular}{l C{0.3cm} C{2cm} C{1.9cm} C{2cm} C{1.9cm} C{2cm}} \hline
\rm Model & & \multicolumn{5}{c}{Best fit} \\
\hline
 & & $\alpha-\Re$ & $\alpha-n$ & $\alpha-\sige$ & $\alpha-\mst^{\rm Chab}$ & $\alpha-\mst$ \\
NFW+light & a b c rms & $-2.288 \pm 0.005$ $0.65 \pm 0.01$ $0.33 \pm 0.02$ $0.148$ & $-1.28 \pm 0.06$ $-1.54 \pm 0.08$ $0.54 \pm 0.03$ $0.261$ & $-1.66 \pm 0.13$ $-1.13 \pm 0.32$ $0.42 \pm 0.18$ $0.274$ & $-2.14 \pm 0.01$ $0.63 \pm 0.06$ $0.35 \pm 0.09$ $0.255$ & $-2.23 \pm 0.01$ $0.35 \pm 0.03$ $0.13 \pm 0.05$ $0.274$ \\
NFW ($\Mvir = 10^{13}\, \rm \Msun$) & a b c rms & $-2.20 \pm 0.01$ $0.64 \pm 0.01$ $0.17 \pm 0.02$ $0.192$ & $-1.29 \pm 0.04$ $-1.48 \pm 0.07$ $0.53 \pm 0.02$ $0.263$ & $-1.11 \pm 0.10$ $-1.97 \pm 0.24$ $0.73 \pm 0.13$ $0.241$ & $-2.12 \pm 0.01$ $0.21 \pm 0.04$ $-0.10 \pm 0.08$ $0.293$ & $-2.17 \pm 0.01$ $0.11 \pm 0.02$ $-0.03 \pm 0.05$ $0.301$ \\
NFW+AC & a b c rms & $-2.19 \pm 0.01$ $0.51 \pm 0.01$ $0.22 \pm 0.02$ $0.138$ & $-1.43 \pm 0.03$ $-1.11 \pm 0.06$ $0.38 \pm 0.02$ $0.217$ & $-1.62 \pm 0.10$ $-0.96 \pm 0.23$ $0.31 \pm 0.13$ $0.214$ & $-2.07 \pm 0.01$ $0.47 \pm 0.03$ $0.25 \pm 0.07$ $0.205$ & $-2.14 \pm 0.01$ $0.25 \pm 0.03$ $0.26 \pm 0.05$ $0.226$ \\
Burkert ($r_{B} = 20 \, \rm kpc$) & a b c rms & $-2.59 \pm 0.01$ $0.24 \pm 0.01$ $0.36 \pm 0.03$ $0.164$ & $-1.29 \pm 0.04$ $-1.57 \pm 0.06$ $0.44 \pm 0.02$ $0.170$ & $-2.34 \pm 0.08$ $-0.43 \pm 0.19$ $0.18 \pm 0.11$ $0.192$ & $-2.54 \pm 0.005$ $0.13 \pm 0.02$ $0.13 \pm 0.05$ $0.197$ & $-2.539 \pm 0.004$ $0.13 \pm 0.01$ $-0.05 \pm 0.02$ $0.195$ \\
Burkert ($r_{B} = 1 \, \rm kpc$) & a b c rms & $-2.615 \pm 0.004$ $-0.10 \pm 0.01$ $-0.12 \pm 0.02$ $0.152$ & $-1.42 \pm 0.04$ $-1.27 \pm 0.05$ $0.30 \pm 0.02$ $0.055$ & $-2.73 \pm 0.12$ $0.12 \pm 0.28$ $-0.03 \pm 0.15$ $0.155$ & $-2.688 \pm 0.005$ $-0.06 \pm 0.02$ $0.16 \pm 0.04$ $0.152$ & $-2.652 \pm 0.004$ $-0.03 \pm 0.01$ $-0.04 \pm 0.02$ $0.158$ \\
NFW (LFS $\cvir-\Mvir$) & a b c rms & $-1.776 \pm 0.005$ $0.55 \pm 0.02$ $-0.35 \pm 0.03$ $0.190$ & $-1.51 \pm 0.03$ $-0.51 \pm 0.05$ $0.19 \pm 0.02$ $0.239$ & $-0.89 \pm 0.17$ $-1.48 \pm 0.40$ $0.40 \pm 0.22$ $0.200$ & $-1.68 \pm 0.01$ $0.32 \pm 0.05$ $0.02 \pm 0.09$ $0.243$ & $-1.8 \pm 0.01$ $0.08 \pm 0.02$ $0.29 \pm 0.04$ $0.253$ \\
S\'ersic light  & a b c rms & $-2.624 \pm 0.005$ $-0.09 \pm 0.01$ $-0.11 \pm 0.02$ $0.141$ & $-1.43 \pm 0.04$ $-1.27 \pm 0.05$ $0.3 \pm 0.02$ $0.055$ & $-2.78 \pm 0.09$ $0.19 \pm 0.23$ $-0.06 \pm 0.13$ $0.158$ & $-2.692 \pm 0.005$ $-0.05 \pm 0.02$ $0.15 \pm 0.04$ $0.152$ & $-2.66 \pm 0.01$ $-0.01 \pm 0.01$ $-0.05 \pm 0.02$ $0.158$ \\
\hline
\end{tabular}
\end{table*}

\begin{table*}
\centering \caption{Best fit parameters with $1\, \sigma$ errors
for the relation \alRe\  $= a + b x + c x^{2}$, where $x = \rm
\log \Re/ \rm 3 kpc$, $n/4$, $\sige/ \rm 200 km/s$, $\rm \log
\mst^{\rm Chab}/(10^{11}\Msun)$ and $\rm \log
\mst/(10^{11}\Msun)$. The scatter (rms) around the best fit
relation is also reported.}\label{tab:tab2}
\begin{tabular}{l C{0.3cm} C{2cm} C{1.9cm} C{2cm} C{1.9cm} C{2cm}} \hline
\rm Model & & \multicolumn{5}{c}{Best fit} \\
\hline
 & & $\alpha-\Re$ & $\alpha-n$ & $\alpha-\sige$ & $\alpha-\mst^{\rm Chab}$ & $\alpha-\mst$ \\
NFW+light & a b c rms & $-2.141 \pm 0.005$ $0.86 \pm 0.01$ $-0.03 \pm 0.02$ $0.170$ & $-2.33 \pm 0.03$ $-0.21 \pm 0.05$ $0.2 \pm 0.01$ $0.292$ & $-1.02 \pm 0.24$ $-2.05 \pm 0.54$ $0.75 \pm 0.30$ $0.341$ & $-1.88 \pm 0.02$ $0.87 \pm 0.03$ $-0.04 \pm 0.07$ $0.303$ & $-2.06 \pm 0.01$ $0.57 \pm 0.02$ $0.23 \pm 0.07$ $0.339$ \\
NFW ($\Mvir = 10^{13} \, \rm \Msun$) & a b c rms & $-1.97 \pm 0.01$ $0.81 \pm 0.01$ $-0.31 \pm 0.02$ $0.224$ & $-2.34 \pm 0.04$ $-0.01 \pm 0.05$ $0.14 \pm 0.02$ $0.285$ & $-0.28 \pm 0.18$ $-3.04 \pm 0.43$ $1.07 \pm 0.23$ $0.292$ & $-1.82 \pm 0.02$ $0.21 \pm 0.04$ $-0.68 \pm 0.13$ $0.348$ & $-1.93 \pm 0.02$ $0.14 \pm 0.03$ $-0.08 \pm 0.09$ $0.359$ \\
NFW+AC & a b c rms & $-2.11 \pm 0.01$ $0.64 \pm 0.01$ $0.04 \pm 0.03$ $0.158$ & $-2.29 \pm 0.03$ $-0.07 \pm 0.04$ $0.12 \pm 0.02$ $0.241$ & $-1.37 \pm 0.13$ $-1.29 \pm 0.31$ $0.42 \pm 0.16$ $0.268$ & $-1.90 \pm 0.01$ $0.68 \pm 0.03$ $0.05 \pm 0.07$ $0.237$ & $-2.03 \pm 0.01$ $0.43 \pm 0.02$ $0.37 \pm 0.05$ $0.272$ \\
Burkert ($r_{B} = 20 \, \rm kpc$) & a b c rms & $-2.6 \pm 0.01$ $0.56 \pm 0.01$ $0.25 \pm 0.03$ $0.205$ & $-2.05 \pm 0.03$ $-1.12 \pm 0.05$ $0.44 \pm 0.02$ $0.265$ & $-2.10 \pm 0.09$ $-1.03 \pm 0.21$ $0.41 \pm 0.10$ $0.305$ & $-2.55 \pm 0.01$ $0.28 \pm 0.02$ $-0.09 \pm 0.05$ $0.311$ & $-2.61 \pm 0.01$ $0.26 \pm 0.02$ $-0.01 \pm 0.04$ $0.321$ \\
Burkert ($r_{B} = 1 \, \rm kpc$) & a b c rms & $-2.832 \pm 0.001$ $-0.062 \pm 0.003$ $-0.037 \pm 0.005$ $0.063$ & $-2.24 \pm 0.02$ $-0.64 \pm 0.03$ $0.15 \pm 0.01$ $0.032$ & $-2.87 \pm 0.05$ $0.03 \pm 0.11$ $-0.01 \pm 0.06$ $0.071$ & $-2.868 \pm 0.002$ $-0.04 \pm 0.01$ $0.09 \pm 0.02$ $0.071$ & $-2.854 \pm 0.001$ $-0.029 \pm 0.005$ $0.02 \pm 0.01$ $0.071$ \\
NFW (LFS $\cvir-\Mvir$) & a b c rms & $-1.716 \pm 0.007$ $0.32 \pm 0.02$ $-0.5 \pm 0.04$ $0.173$ & $-2.1 \pm 0.04$ $0.38 \pm 0.06$ $-0.09 \pm 0.02$ $0.195$ & $-1.64 \pm 0.27$ $-0.04 \pm 0.66$ $-0.22 \pm 0.38$ $0.192$ & $-1.673 \pm 0.005$ $0.38 \pm 0.02$ $0.21 \pm 0.05$ $0.184$ & $-1.76 \pm 0.01$ $0.19 \pm 0.02$ $0.34 \pm 0.03$ $0.214$\\
S\'ersic light  & a b c rms & $-2.841 \pm 0.002$ $-0.041 \pm 0.005$ $-0.05 \pm 0.01$ $0.063$ & $-2.25 \pm 0.02$ $-0.64 \pm 0.03$ $0.15 \pm 0.01$ $0.032$ & $-2.91 \pm 0.05$ $0.08 \pm 0.11$ $-0.02 \pm 0.06$ $0.071$ & $-2.871 \pm 0.002$ $-0.02 \pm 0.01$ $0.066 \pm 0.02$ $0.071$ & $-2.856 \pm 0.002$ $-0.01 \pm 0.01$ $-0.02 \pm 0.01$ $0.071$ \\
\hline
\end{tabular}
\end{table*}

These results extend our previous analysis
in~\cite{TRN13_SPIDER_IMF}, in that we explore here a larger set
of DM models (i.e. Burkert and ``high-concentration'' NFW models),
and present also the correlations of \dimf\ with \Re, $n$, and
mass (besides that with $\sigma$), and are consistent with a
plethora of independent results from dynamical and stellar
population studies (\citealt{Treu+10}; \citealt{ThomasJ+11};
\citealt{Conroy_vanDokkum12b}; \citealt{Cappellari+12,
Cappellari+13_ATLAS3D_XX}; \citealt{Spiniello+12};
\citealt{Wegner+12}; \citealt{Dutton+13}; \citealt{Ferreras+13};
\citealt{Goudfrooij_Kruijssen13};
\citealt{LaBarbera+13_SPIDERVIII_IMF}; \citealt{TRN13_SPIDER_IMF};
\citealt{Weidner+13_giant_ell}; \citealt{Tortora+14_MOND};
\citealt{Goudfrooij_Kruijssen14}; \citealt{Shu+14_SLACSXII}). We
notice that for ``high-concentration'' NFW models the best-fitting
\Yst\ is significantly lower (by $ \sim 0.5$~dex, at $\sigma_{\rm
e}$ $\sim 100$~$\rm km \, s^{-1}$) than that for  a Chabrier IMF
(i.e. \dimf $< 1$). Since the Chabrier IMF gives a minimum
normalization with respect to either top- or bottom-heavier
distributions (see above), our data seem to be more consistent
with a somewhat lower concentration than that of the LFS12 \cvir
-- \Mvir\ relation (see Section~\ref{sec:analysis}), although
measurement uncertainties on \cvir\ and $\Yst^{\rm Chab}$ might
indeed be responsible for the low ($<1$) \dimf\ values.

\section{Total mass density slopes}\label{sec:results}

\subsection{Correlations with galaxy properties}

We start by presenting results for the SPIDER sample, where the
stellar mass profiles of ETGs are characterized with S\'ersic fits
of NIR (K-band) galaxy images (see Sec.~\ref{sec:sample_SPIDER}).
In this section, we focus the discussion on results for our
reference NFW model, comparing those for different models in
Section~\ref{sec:comp_models}. Fig.~\ref{fig:slopes_1} shows the
correlations of logarithmic mass slopes, \alRetwo\ (top) and
\alRe\ (bottom), as a function of  \Re, S\'ersic $n$, \sige , and
stellar mass (using either a Chabrier IMF, or the IMF
normalization provided by our best-fit model for each galaxy). The
\sige\ is the SDSS-fibre velocity dispersion, $\rm \sigAp$,
corrected to an aperture of one \Re , following
\cite{Cappellari+06}. Fig.~\ref{fig:slopes_2} shows the same
correlations as in Fig.~\ref{fig:slopes_1} but for the
mass-weighted (rather than local) slope, \amwRe . Comparing the
Figures, one can see that the slope value depends significantly on
its definition, i.e. \alRetwo\ $\neq$ \alRe\ $\neq $~\amwRe\
within the uncertainties, suggesting that either the total mass
profile of ETGs is not exactly a power-law, or the explored set of
models -- which are non-power-laws by construction -- is not able
to describe accurately a power-law behavior of the profiles (see
Sec.~\ref{sec:def_slopes}). More (spatially extended) kinematical
data would be necessary to address this issue.

In Tables~\ref{tab:tab1}, \ref{tab:tab2} and~\ref{tab:tab3} we
show the results of fitting the trends in Figs.~\ref{fig:slopes_1}
and~\ref{fig:slopes_2}, with $2^{\rm nd}$ order polynomials of the
form $\alpha = a + b x + c x^{2}$. Errors on slopes are computed
by a bootstrap method, and are quoted at the $1\rm \sigma$ level.
Almost all the correlations are significant at more than $99\%$.

For our reference, NFW, DM models (solid black curves in the
Figures), the \al\ becomes shallower with galaxy mass and radius,
reaching, for the highest radii probed, a value of about $-1.5$,
i.e. even shallower than the isothermal value ($-2$). Milder
trends of \al , than those for \Re\ and mass, are observed with
respect to S\'ersic $n$ and \sige . At $\Re/2$, the \al\ exhibits
a double-value behaviour as a function of $n$, increasing at both
high and low $n$, while \alRe\ tends to steepen with $n$. In terms
of \sige , both \alRetwo\ and \alRe\ tend to steepen with velocity
dispersion. For \amwRe , as it might be expected, the trends are
intermediate between those for \alRetwo\ and \alRe . In general,
\amwRe\ increases with mass and radius, while it exhibits a
double-value behavior with $n$, and mildly decreases with \sige ,
consistent with the findings for \alRetwo\ and \alRe.

Figs.~\ref{fig:slopes_1} and~\ref{fig:slopes_2} also plot the
slopes of the stellar mass profile only (shaded regions), obtained
from the best-fitting K-band S\'ersic profiles, under the
assumption of a radially-constant (stellar) \ML\ (see
Sec.~\ref{sec:analysis}). In contrast to \al, the stellar mass
slope does not vary significantly with \Re\ and \mst . No
significant variation with \sige\ is observed (similar to \al),
while the stellar mass slope tends to steepen with $n$, as
expected by the fact that as $n$ increases the shape of the
S\'ersic law becomes more peaked towards the {centre}.
Interestingly, at low \Re\ and \mst , the NFW-based total mass
density slope approaches the slope of the stellar mass component,
i.e. that for a constant-\ML\ profile. {This is due to the fact
that in the centre of low-mass (small) ETGs, the stellar mass
distribution dominates the total mass budget.} Notice that
\cite{Tortora+09} reached a similar conclusion by comparing
central DM density estimates with predictions of $\Lambda$CDM
toy-models.

\begin{table*}
\centering \caption{Best fit parameters with $1\, \sigma$ errors
for the relation $\amwRe = a + b x + c x^{2}$, where $x = \rm \log
\Re/ \rm 3 kpc$, $n/4$, $\sige/ \rm 200 km/s$, $\rm \log \mst^{\rm
Chab}/(10^{11}\Msun)$ and $\rm \log \mst/(10^{11}\Msun)$. The rms
is also reported. The scatter (rms) around the best fit relation
is also reported.}\label{tab:tab3}
\begin{tabular}{l C{0.3cm} C{2cm} C{1.9cm} C{2cm} C{1.9cm} C{2cm}} \hline
\rm Model & & \multicolumn{5}{c}{Best fit} \\
\hline
 & & $\alpha-\Re$ & $\alpha-n$ & $\alpha-\sige$ & $\alpha-\mst^{\rm Chab}$ & $\alpha-\mst$ \\
NFW & a b c rms & $-2.218 \pm 0.004$ $0.50 \pm 0.01$ $0.28 \pm 0.01$ $0.122$ & $-1.47 \pm 0.04$ $-1.12 \pm 0.06$ $0.38 \pm 0.02$ $0.212$ & $-1.73 \pm 0.09$ $-0.85 \pm 0.2$ $0.31 \pm 0.10$ $0.212$ & $-2.1 \pm 0.01$ $0.51 \pm 0.3$ $0.35 \pm 0.06$ $0.195$ & $-2.16 \pm 0.01$ $0.28 \pm 0.02$ $0.09 \pm 0.04$ $0.210$ \\
NFW ($\Mvir = 10^{13} \, \rm \Msun$) & a b c rms & $-2.139 \pm 0.005$ $0.48 \pm 0.01$ $0.13 \pm 0.02$ $0.161$ & $-1.45 \pm 0.04$ $-1.08 \pm 0.06$ $0.38 \pm 0.02$ $0.217$ & $-1.24 \pm 0.09$ $-1.62 \pm 0.19$ $0.61 \pm 0.10$ $0.184$ & $-2.08 \pm 0.01$ $0.14 \pm 0.03$ $-0.08 \pm 0.06$ $0.235$ & $-2.11 \pm 0.01$ $0.07 \pm 0.02$ $-0.07 \pm 0.04$ $0.239$ \\
NFW+AC & a b c & $-2.147 \pm 0.005$ $0.40 \pm 0.01$ $0.20 \pm 0.02$ $0.130$ & $-1.57 \pm 0.03$ $-0.83 \pm 0.04$ $0.27 \pm 0.01$ $0.190$ & $-1.62 \pm 0.13$ $-0.93 \pm 0.32$ $0.35 \pm 0.18$ $0.184$ & $-2.05 \pm 0.01$ $0.41 \pm 0.04$ $0.25 \pm 0.08$ $0.176$ & $-2.11 \pm 0.01$ $0.21 \pm 0.02$ $0.23 \pm 0.03$ $0.192$ \\
Burkert ($r_{B} = 20 \, \rm kpc$) & a b c rms & $-2.431 \pm 0.004$ $0.13 \pm 0.01$ $0.16 \pm 0.02$ $0.122$ & $-1.51 \pm 0.02$ $-1.1 \pm 0.04$ $0.30 \pm 0.01$ $0.105$ & $-2.25 \pm 0.08$ $-0.31 \pm 0.19$ $0.13 \pm 0.11$ $0.130$ & $-2.401 \pm 0.004$ $0.07 \pm 0.01$ $0.05 \pm 0.03$ $0.134$ & $-2.398 \pm 0.004$ $0.09 \pm 0.01$ $-0.07 \pm 0.01$ $0.134$ \\
Burkert ($r_{B} = 1 \, \rm kpc$) & a b c rms & $-2.463 \pm 0.004$ $-0.09 \pm 0.01$ $-0.1 \pm 0.01$ $0.110$ & $-1.58 \pm 0.02$ $-0.92 \pm 0.03$ $0.21 \pm 0.01$ $0.032$ & $-2.58 \pm 0.09$ $0.17 \pm 0.22$ $-0.05 \pm 0.13$ $0.126$ & $-2.524 \pm 0.004$ $-0.05 \pm 0.02$ $0.14 \pm 0.04$ $0.118$ & $-2.49 \pm 0.01$ $-0.02 \pm 0.01$ $-0.05 \pm 0.02$ $0.126$ \\
NFW (LFS $\cvir-\Mvir$) & a b c rms & $-1.804 \pm 0.004$ $0.44 \pm 0.01$ $-0.27 \pm 0.02$ $0.164$ & $-1.55 \pm 0.03$ $-0.42 \pm 0.04$ $0.15 \pm 0.01$ $0.205$& $-1.06 \pm 0.12$ $-1.29 \pm 0.28$ $0.39 \pm 0.16$ $0.164$ & $-1.72 \pm 0.01$ $0.29 \pm 0.04$ $0.07 \pm 0.08$ $0.20$ & $-1.83 \pm 0.01$ $0.08 \pm 0.02$ $0.27 \pm 0.03$ $0.210$\\
S\'ersic light  & a b c rms & $-2.469 \pm 0.004$ $-0.08 \pm 0.01$ $-0.09 \pm 0.01$ $0.114$ & $-1.60 \pm 0.02$ $-0.90 \pm 0.03$ $0.2 \pm 0.01$ $0.032$ & $-2.60 \pm 0.08$ $0.18 \pm 0.20$ $-0.06 \pm 0.12$ $0.126$ & $-2.53 \pm 0.004$ $-0.04 \pm 0.02$ $0.14 \pm 0.04$ $0.122$ & $-2.497 \pm 0.003$ $-0.01 \pm 0.01$ $-0.04 \pm 0.02$ $0.126$ \\
\hline
\end{tabular}
\end{table*}

For the \atlas3d\ sample, we get, in general, consistent results
with those for the SPIDER sample. Fig.~\ref{fig:SPIDER_vs_A3D}
(left-panel) compares, for example, the trends of \alRe\ with
\mst\ for both samples.  The best fit trend for SPIDER shown in
Table \ref{tab:tab2} is \alRe\ $= -1.88 + 0.87\mst - 0.04\mst^{2}$
with scatter of $rms = 0.303$, while for \atlas3d\ we find \alRe\
$= -2.23 + 0.68\mst + 0.69\mst^{2}$ and a scatter of $rms =
0.207$. We remark that SPIDER and \atlas3d\ ETGs are analyzed here
with the same approach, although the galaxy light distributions
and stellar masses are characterized in significantly different
ways.   In fact, the $n=4$ light profiles for \atlas3d\ galaxies
have shallower slopes with respect to the SPIDER S\'ersic laws
\footnote{ \cite{Cappellari+13_ATLAS3D_XV} found that stellar
light profiles were well fitted by an isothermal law within 1 \Re
(see their Fig. 2). Our stellar-light slopes for the \atlas3d\
galaxies would be consistent with their findings if the same slope
definition were adopted.}. The agreement between the two data-sets
is good, although the trend with mass for \atlas3d\ tends to be
shallower than that for SPIDER, with steeper slopes at high
masses, because of the shallower (de Vaucouleurs vs. S\'ersic)
light profile. At the lowest \mst\ ($\sim 10^{10} \, M_\odot$),
which can be probed only with \atlas3d, one can observe an
inversion of the mass density trend with the slope, with \al\
becoming shallower than at \mst $\sim 10^{10.5} \, M_\odot$.
However, this result might be just reflecting the fact that the
$r^{1/4}$ law is not very accurate for low, relative to high, mass
ETGs. The right panel of Fig.~\ref{fig:SPIDER_vs_A3D} also shows
the correlation of mass density slopes with central DM fraction
within \Re\, $\fdm (\Re)$. As for best fits in
Tables~\ref{tab:tab1}, \ref{tab:tab2} and~\ref{tab:tab3}, the
trend is quite well fitted by a polynomial. We find \alRe\ $=
-2.65 + 3.42 x -2.52 x^{2}$, with $x = \fdm(\Re)$ and a scatter of
0.07. For \atlas3d\ the fit is \alRe\ $= -2.73 + 3.91 x -3.94
x^{2}$ with the same scatter found for SPIDER sample. We find
consistent results between the two data-sets, with shallower
density profiles in galaxies that are more DM dominated in the
center, consistent with independent results
from~\citet{Auger+10_SLACSX} and~\cite{Dutton_Treu14} (see below).

Fig.~\ref{fig:SPIDER_vs_A3D} also compares the SPIDER and
\atlas3d\ trends with those obtained for SPIDER ETGs, by computing
density mass slopes and dark matter fractions with $r^{1/4}$ (i.e.
de Vaucouleurs) structural parameters in r band from SDSS-DR6,
rather than Sersic 2DPHOT parameters in K band (see
\cite{SPIDER-I} for details). This comparison allows us to single
out the effect of differences due to light profile shape, from
those of different wavebands and sample selection, on the observed
trends. The $r^{1/4}$ trends with \mst\ are shallower than
reference ones for SPIDER, and fairly consistent with those for
\atlas3d, the small difference between solid (\atlas3d ) and
dashed (SPIDER $r^{1/4}$) gray curves being likely explained by
differences in sample selection. The fact that $r^{1/4}$, with
respect to S\'ersic, parameters provide steeper density slopes is
also consistent with the trends in the bottom panels of
Fig.~\ref{fig:slopes_1}. In fact, fitting a $n=4$ light profile
gives smaller \Re\ values than those for a S\'ersic law, with
steepest $\alpha$ values being found for $n \sim 4$ and the
smallest \Re .}

\subsection{Comparison of different DM models}

\label{sec:comp_models} We discuss here how different assumptions
on the DM component affects the trends of the mass density slope.
As shown in Figs.~\ref{fig:slopes_1} and~\ref{fig:slopes_2}, the
slope values are degenerate with halo model (see also best fits in
Tables~\ref{tab:tab1}, \ref{tab:tab2} and~\ref{tab:tab3}). For
most correlations, the Burkert and ``high-concentration'' NFW
models (see blue and green curves) provide slope values
encompassing the whole range of values for \al , with the
reference NFW model being in between these models (\citealt{CT10};
\citealt{Cardone+11SIM}). { Notice that the estimate of total mass
density slope is deeply related to the best-fitting \Yst, as for
increasing \Yst\ the mass budget in the central regions resembles
more the one for the light component alone. In fact,} contracted
halo models, which imply a larger DM content towards the galaxy
center, with smaller \Yst\ (see Fig.~\ref{fig:dIMF}), tend to give
shallower slopes than the reference NFW models, with this behavior
being even more pronounced for ``high-concentration'' models. On
the other hand, Burkert profiles provide steeper slopes, in
between those for NFW models and stellar mass density (i.e.
constant-\ML) slopes (\citealt{TRN13_SPIDER_IMF}). Remarkably, in
the case of Burkert models, the slopes show almost constant trends
with all galaxy properties, including mass and radius, in sharp
contrast with the significant trends obtained for all other
models. In particular, the results for the model with $r_{\rm B} =
1 \, \rm kpc$ closely resemble the slopes of the stellar mass
distribution. Notice that, different than for NFW models, we have
not adopted a trend of core radius with galaxy mass (equivalent to
an \Mvir--\mst\ relation) for Burkert models, but just two
reference values of $\rm r_{\rm B}=1$ and $20$~kpc. These values
approximately bracket the results found for two elliptical
galaxies by MSB11 and the range of core radii obtained by
\cite{ThomasJ+09} using cored logarithmic haloes, which resemble
Burkert profiles in the galactic centers. Thus, using a
radius-mass relation would not change significantly our results
for the Burkert profiles.

Notice that NFW models with fixed virial mass and concentration
(red curves in the Figures) give shallower slopes than, but
similar trends to, the NFW case. In some cases, the slopes are
also (marginally) shallower than those for NFW contracted
profiles. For ``high-concentration'' models, the trends deviate
significantly -- with higher (i.e. shallower) $\alpha$ values --
from our reference model.   This is more pronounced at low-
relative to high-mass, making the trends of $\alpha$ with mass
significantly shallower than for reference models. On the other
hand, the trends with radius and \sige\ are more robust to the
\cvir -- \Mvir\ prescription, in particular for \alRetwo .

In summary, we find that the strong increase of the mass density
slope with galaxy radius, as well as the decrease with \sige , are
robust findings against different ingredients of NFW halo models.
The trend with mass is less robust, in that it is significantly
shallower for ``high-concentration'' models. However, as noticed
in Sec.~\ref{sec:IMF_trends}, the ``high-concentration'' models
provide overly low IMF normalizations at low galaxy mass (i.e.
lower than those measured for a Chabrier IMF), which might favour
(somewhat) lower concentration profiles. {Moreover, one can notice
that the \cvir--\Mvir\ relation from LFS12 is derived by assuming
a fixed Chabrier IMF. Although this might be important at high
galaxy mass, where the IMF normalization is found to be larger
than the Chabrier one (e.g. \citealt{TRN13_SPIDER_IMF}, and
Fig.~\ref{fig:dIMF}), we find fair agreement, at high radius/large
mass, between density slopes for fiducial NFW and
``high-concentration'' models.} In contrast, all correlations tend
to be washed out when using Burkert profiles. However, such
models, while reproducing quite well the dynamics of dwarf
galaxies and spirals, likely provide too \emph{light} haloes in
massive ETGs (\citealt{CT10}), with respect to many results
pointing to a significant amount of DM at the virial radius in
these systems (e.g., \citealt{Benson+00}; \citealt{MH02};
\citealt{vdB+07}; \citealt{Moster+10}). Therefore, while we have
included here also Burkert models in the analysis, these should be
considered as rather unlikely for the most massive galaxies in our
samples.

\subsection{Impact of different assumptions}

Our dynamical approach relies on several assumptions, i.e. (i)
spherical symmetry, (ii) isotropy, (iii) no stellar \ML-gradients
within a galaxy, and (iv) no significant rotation. We have
performed a variety of tests, showing that these assumptions do
not affect significantly the correlations of mass density slope
with galaxy properties.

\begin{description}
 \item[(i)] In general, for a flattened system, the use of spherical
models tends to over- (under-) estimate the inferred galaxy mass,
if the system is seen edge-on (face-on). To minimize the fraction
of non-spherical systems (e.g. S0's), for both SPIDER and
\atlas3d\ samples, we have restricted the analysis to ``round''
objects, with axis ratio $q> 0.8$. We found that the slope trends
remain unchanged, within a few percent, with respect to those for
the entire samples.
 \item[ii)] To explore the effect of radial anisotropy, we have adopted two
empirically motivated values of the radial anisotropy parameter,
$\beta=+0.1$ and $+0.2$, respectively (\citealt{Gerhard+01}). For
$\beta>0$, the model velocity dispersion at a given radius is
larger than for $\beta=0$, with the net effect of reducing  our
inferred masses within that radius. For $\beta=+0.1$ and $+0.2$,
the inferred masses at $1$~\Re\ were found to change by $\sim 2$
and $4$~\%, respectively, with negligible impact on the mass
density slopes, considering the observed scatter of slope values.
 \item[iii)] Radial gradients of \ML\ can also affect our density slope estimates. However,
at optical wavebands, such gradients are very mild in massive ETGs
(\citealt{Tortora+11MtoLgrad}), and are expected to be even
smaller in the NIR, where most of the integrated light from a
stellar population is dominated by its old  quiescent component.
Indeed, the fact that for r-(\atlas3d) and K-(SPIDER)band data, we
find consistent slope estimates, indicates that \ML\ gradients are
not important for our analysis.
\item[iv)] The \atlas3d\ sample gives us the opportunity to test
the impact of neglecting rotational velocity and galaxy flattening
in our analysis. From best-fitting JAM models, the \atlas3d team
obtained the best-fitting relation $V_{\rm circ}^{\rm obs}(R_{\rm
e, maj}) \approx 1.51 \times \sige$, where $V_{\rm circ}$ is the
model circular velocity and $R_{\rm e, maj}$ is the effective
radius along the galaxy major axis
(\citealt{Cappellari+13_ATLAS3D_XV}). Using the expression $V_{\rm
circ}^{\rm theo} \equiv \sqrt{G \Mdyn / r}$ we have converted the
$V_{\rm circ}^{\rm obs}(R_{\rm e, maj})$ from \atlas3d\ to a
dynamical mass, \Mdyn. Even in this case, we found that neglecting
rotation has negligible effects, within a few percent, on the mass
density slopes.
\end{description}

\begin{figure*}
\psfig{file=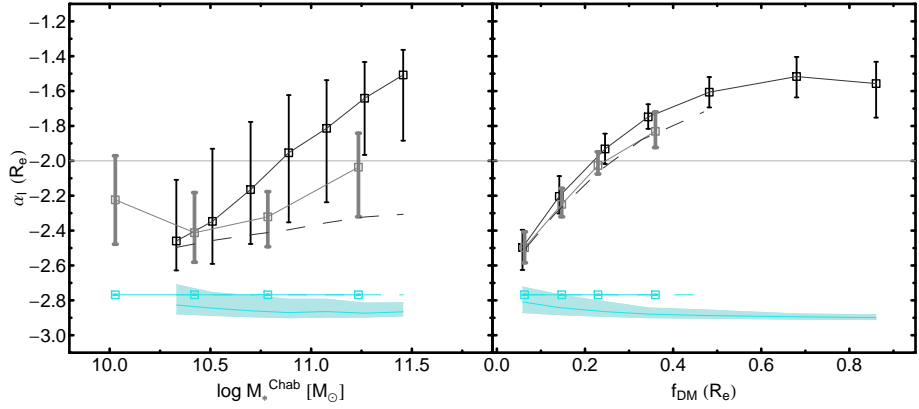, width=0.7\textwidth} \caption{Comparison of
trends of mass density slope, \alRe , with \mst\ (left) and DM
fraction, $\fdm(\Re)$ (right), between SPIDER (black squares) and
\atlas3d\ (gray squares) samples. Error bars mark the 16-84th
percentile scatter intervals on the slopes. { Cyan shaded regions
and squares mark the stellar mass slopes, at one \Re , for SPIDER
and \atlas3d , respectively.} Notice that for \atlas3d, the slope
of the light distribution is constant, as the light profile is
parameterized by a (fixed-shape) de Vaucouleurs model. { As a
comparison, we also plot, as dashed curves, the trends obtained
for SPIDER ETGs when using SDSS-DR6 r-band de Vaucouleurs (rather
than K-band S\'ersic) models to parameterize the galaxy light
profiles (dashed lines).}}\label{fig:SPIDER_vs_A3D}
\end{figure*}

\section{Comparison with literature}\label{sec:literature}

\subsection{Observations}\label{sec:observations}

Fig.~\ref{fig:SPIDER_vs_literature} compares some of our findings with
independent estimates of the mass density slope from the literature. At
the highest mass scales  probed in the present work, our results
are consistent with \citet{Auger+10_SLACSX}, who fitted a sample
of SLACS lenses, at intermediate redshifts, with a power-law mass
density profile, $\rho(r) \propto r^{\alpha}$, combining
gravitational lensing and central dynamics to probe the total
mass distribution at $\Re/2$. They found the mass distribution to
be well described by an isothermal profile (\citealt{TK04};
\citealt{Gavazzi+07_SLACSIV}; \citealt{Auger+10_SLACSX}). The
average slope value from~\cite{Auger+10_SLACSX} is plotted in
the top panels of Fig.~\ref{fig:SPIDER_vs_literature} (see red squares and error bars),
vs. \mst\ (left) and $\fdm(\Re/2)$ (right), respectively. Notice
that \citet{Auger+10_SLACSX} derived stellar masses by assuming a
Salpeter IMF. Therefore, to perform a meaningful comparison, we also converted
our Chabrier-IMF \mst's into Salpeter-IMF stellar masses,
accounting for the different overall normalizations of the
Chabrier and Salpeter IMFs.

\begin{figure*}
\psfig{file=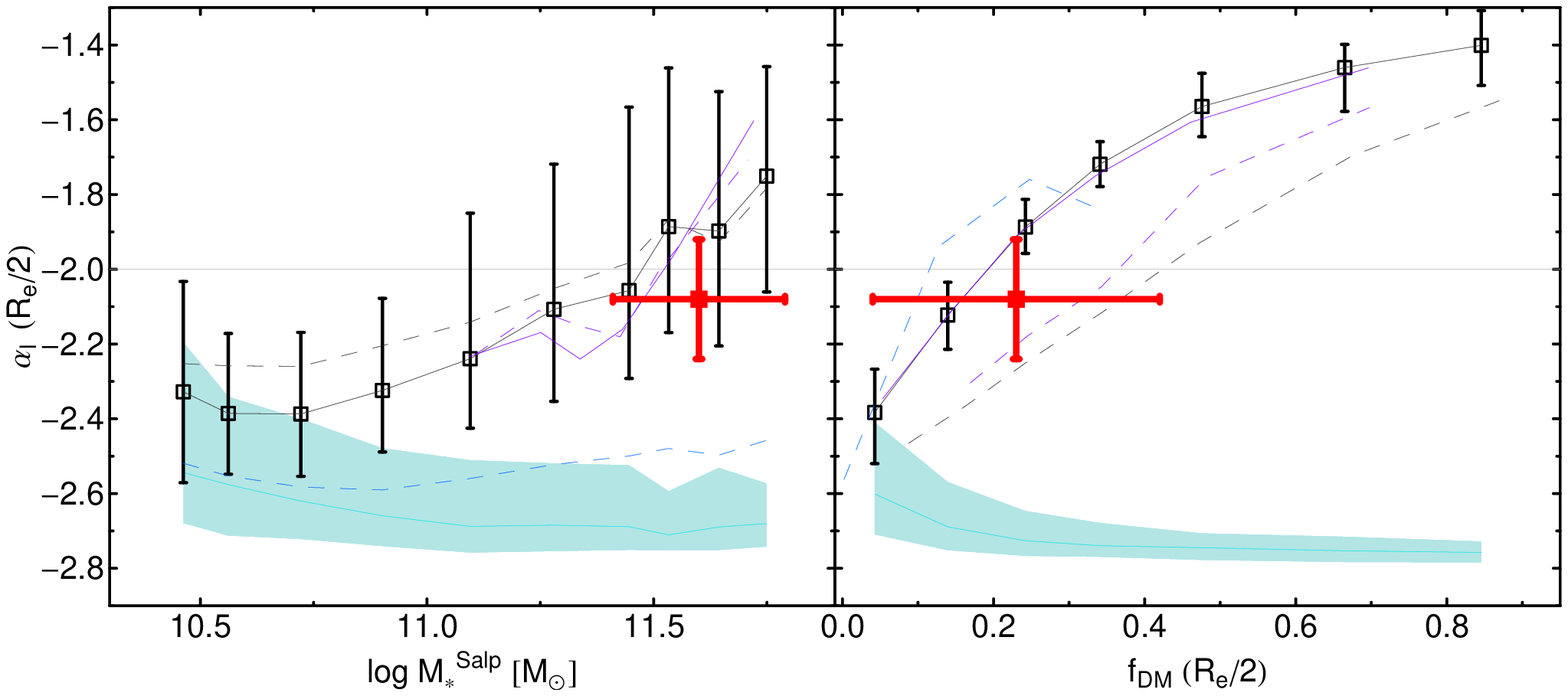, width=0.7\textwidth}\\
\psfig{file=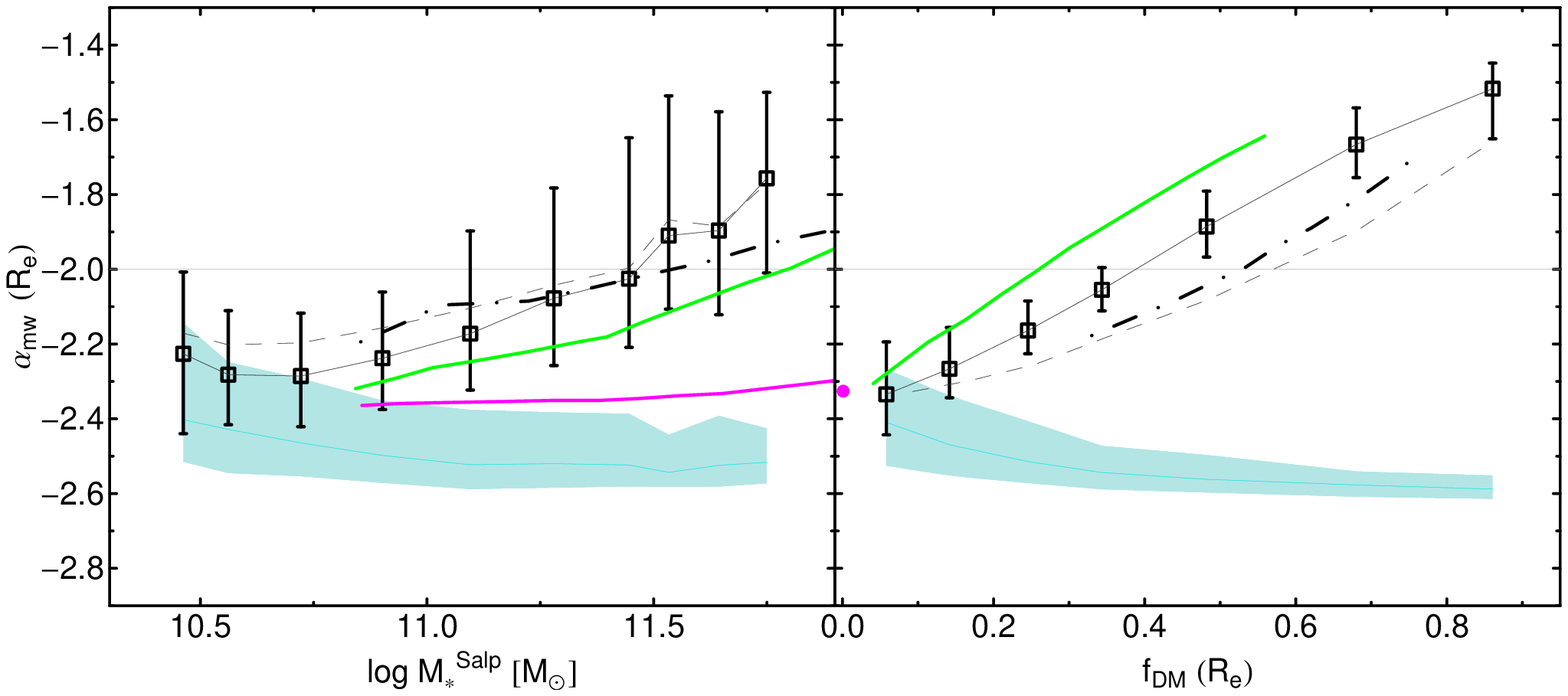, width=0.7\textwidth} \caption{Comparison of
total mass density profile slopes with data from the literature,
as a function of \mst\ (left panels) and  \fdm\ (right panels).
Black symbols with error bars and cyan lines and regions are the
same as in Figs.~\ref{fig:slopes_1} and~\ref{fig:slopes_2},
plotting the slope trends of SPIDER ETGs, for our fiducial,
NFW+S\'ersic, model (black curve and error bars) and the slopes
for the stellar mass profile only (cyan curve and shaded region),
respectively. Top and bottom panels are for \alRetwo\ and \amwRe ,
respectively. Literature data include: average slope and $1 \,
\sigma$ scatter for SLACS lenses~\citep{Auger+10_SLACSX}, plotted
with red squares and error bars in the top panels; slopes obtained
for constant-\ML\ (magenta curves), NFW (green curves) and
contracted-NFW (dot dashed black curves) profiles from
\citet{Dutton_Treu14} in the bottom panels. Purple solid and
dashed lines in the top panels are the slope trends for SPIDER
ETGs with Salpeter IMF normalization (see the text for details).
}\label{fig:SPIDER_vs_literature}
\end{figure*}

The agreement between~\citet{Auger+10_SLACSX} -- who estimated the
density slope of massive early-type lenses -- and our most massive
bin for fiducial NFW models, is excellent, with a good agreement
also with respect to \fdm . The best fitted relation
\alRetwo--$\mst^{\rm Chab}$ shown in Table \ref{tab:tab1}, because
of the change of IMF, is now \alRetwo\ $= -2.28 + 0.40 x +0.40
x^{2}$, with $x = \mst^{\rm Salp}$ and a scatter of $rms=0.255$.
The trend with \fdm\ is best fitted by the polynomial \alRetwo\ $=
-2.49 + 2.86 x -1.91 x^{2}$, with $x = \fdm(\Re/2)$ and a scatter
of $rms=0.130$.

However, \citet{Auger+10_SLACSX} assumed a fixed, Salpeter, IMF,
while our dynamical approach leaves the IMF normalization --
through the best-fitting stellar mass-to-light ratio -- as a free
model parameter. To test the effect of IMF normalization, we have
selected only galaxies in our sample that are best described by a
Salpeter-like normalization (with $1.6 < \dimf <2$). This
selection leads to mild variations ($<10 \%$) in the slope trends
at high galaxy mass (see purple curves in the Figure), with slopes
still in excellent agreement with SLACS. The agreement is good
also when we compare the trends with \Re\ and  velocity dispersion
with ours in the top panels in Fig.~\ref{fig:slopes_1}, as
\cite{Auger+10_SLACSX} find shallower slopes at larger \Re\ and an
almost constant trend with velocity dispersion.

Fig.~\ref{fig:SPIDER_vs_literature} also shows that Burkert
profiles (dashed and solid blue curves in the top panels) give
slopes that are too steep (at the $2.5$~$\sigma$ level) with
respect to SLACS. Thus, the comparison of our dynamical analysis
with gravitational lensing results at intermediate redshift seems
to reject Burkert profiles as plausible models to describe the DM
component of (massive) ETGs, while it is fully consistent with
massive ETGs having an isothermal total mass density profile.

In the bottom panels of Fig.~\ref{fig:SPIDER_vs_literature} we
compare our findings, in terms of \amwRe , with the dynamical
analysis performed for a SDSS sample of ETGs, by \citet[hereafter
DT14]{Dutton_Treu14}. The \amwRe\ is plotted vs. \mst\ (left) and
$\fdm(\Re)$ (right). The best fitted relation \alRetwo--$\mst^{\rm
Salp}$ is \amwRe\ $= -2.21 + 0.31 x +0.38 x^{2}$, with $x =
\mst^{\rm Salp}$ and a scatter of $rms=0.196$. The trend with
\fdm\ is best fitted by the polynomial \amwRe\ $= -2.41 + 1.06 x
-0.01 x^{2}$, with $x = \fdm(\Re)$ and a scatter of $rms=0.095$.
The authors modelled the galaxy light profiles with the
combination of an $n = 1$ and an $n = 4$ S\'ersic law, with a
suite of models to describe the DM distribution (including
fiducial NFW, constant-\ML , contracted, and expanded models),
with varying stellar \ML . The Figure compares their findings with
ours, for NFW, contracted-NFW, and constant-\ML\ profiles. In
general, the shape of the correlations are similar when comparing
ours and DT14 results, but some offsets, at the $10 \%$ level,
exist. In particular, constant-\ML\ models from DT14 (magenta
curves) have slopes $\sim 8\%$ shallower than ours (cyan curve and
shaded region), while mass density slopes for NFW models are
steeper (shallower) than ours when plotted with respect to \mst\
(\fdm). A good agreement is found for contracted-NFW models.  {
The offset between our NFW-model slopes and those of DT14 is
likely due to the different modeling of the galaxy light
distribution between our study and theirs. In fact, as shown in
Fig.~\ref{fig:slopes_2}, the \amwRe\ of  SPIDER ETGs with $n \sim
4$ is lower (i.e. steeper) than that for both higher- and
lower-$n$ galaxies, especially for NFW-model slopes, suggesting
that a combination of $n=1$ and $n=4$ S\'ersic laws, to model the
light distribution of ETGs, can produce lower (steeper) \amwRe\
slopes than those for a single S\'ersic law with variable $n$,
consistent with what seen in the comparison of DT14 and our
trends. A good agreement is found for the trend with \Re, while
DT14 find shallower slopes in high-\sigs\ galaxies, in agreement
with recent findings from gravitational lenses
(\citealt{Shu+14_SLACSXII}), but in contrast with our constant
trends (see Fig.~\ref{fig:slopes_2}).

The fact that the mass density slope becomes shallower at high-,
relative to low-, mass is also consistent with the findings
of~\citet[hereafter HB10]{Humphrey_Buote10}. Using {\it Chandra}
X-ray data, under the assumption of hydrostatic equilibrium, HB10 analyzed a sample
of 10 systems, spanning $\sim 2$ orders of magnitude in \Mvir , from
massive galaxies to clusters of galaxies, in the radial range
from $\lsim \Re$ to several \Re's. They found isothermal profiles
for galaxies, consistent with our results for massive ETGs, and
shallower than isothermal slopes (up to $\alpha = -1.2$) for galaxy clusters.

\begin{figure*}
\psfig{file=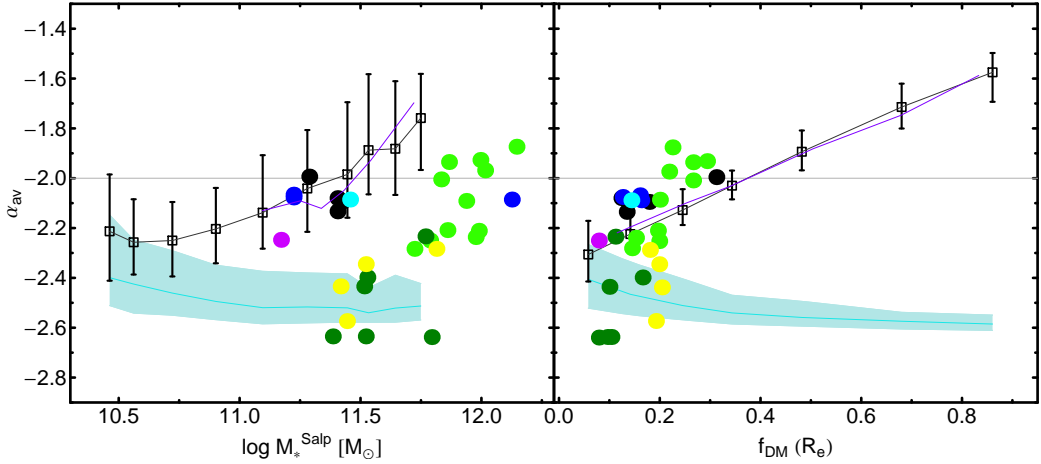, width=0.8\textwidth} \caption{Comparison of
total mass density slope trends with predictions for simulated
galaxies from~R13. Left and right panels plot $\alpha_{\rm av}$
vs. \mst\ and \fdm\ (within $1$~\Re ), respectively. { The slope,
$\alpha_{\rm av}$, is defined by fitting mass density profiles
with a power-law, in the radial range $0.3$ -- $4 \, r_{1/2}$,
where $r_{1/2}$ is the galaxy half-mass radius (i.e. adopting the
same slope definition as in R13, see the text for details).} Black
symbols with error bars and cyan lines and regions are the same as
in Figs.~\ref{fig:slopes_1} and~\ref{fig:slopes_2}. Our stellar
masses are re-scaled to a Salpeter IMF, as in R13. Purple lines
are the observed trends for the subsample of SPIDER ETGs with
Salpeter-like IMF normalization (see Sec.~\ref{sec:observations}).
Dots with different colours are simulated galaxies from~R13:
black, blue, cyan and pink dots correspond to idealized single
binary mergers, while light-green, dark-green, and yellow dots are
for mergings systems drawn from cosmological simulations (see
Sec.~\ref{sec:simulations} for details). Notice the good agreement
of our trends with binary mergers simulations, including also the
effect of BH growth and feedback.}\label{fig:simulations}
\end{figure*}

\subsection{Simulations}
\label{sec:simulations}

We compare our results with predictions for a suite of simulated
galaxies, from \citet[hereafter R13]{Remus+13}. The comparison is
shown in Fig.~\ref{fig:simulations}, where we plot mass density
slopes, for our fiducial (NFW+S\'ersic) model, as a function of
stellar mass (left) and central DM  fraction (right). R13 computed
mass density slopes by fitting mass density profiles with a
power-law, in the radial range $0.3$ -- $4 \, r_{1/2}$, where
$r_{1/2}$ is the half-mass radius of a simulated galaxy. To
perform a meaningful comparison, we re-computed our slopes with
the same definition as in R13, converting the projected effective
radius of a given galaxy into its half-mass (de-projected) radius.
We used {the relation $\Re = r_{1/2} / 1.35$, which turns out to
be independent of the S\'ersic $n$ (see, e.g., appendix~B of
\citealt{Wolf+10}).} We refer to the slopes, defined as in R13, as
$\alpha_{\rm av}$.

Fig.~\ref{fig:simulations} plots the $\alpha_{\rm av}$ for all
models from R13 (see dots with different colours), except for
simulated BCGs, whose mass range is above that covered by our
trends. The simulations include several high-resolution ``binary
mergers'', i.e. a) spiral-spiral mergers with a progenitor mass
ratio $1:1$ (black), b) spiral-spiral mergers with mass ratio
$3:1$ (blue), c) a mixed merger of a spiral galaxy with an
elliptical, the latter formed by a 3:1 spiral-spiral merger
(cyan), d) one spiral-spiral merger, with a mass ratio $3:1$, and
a large gas fraction of 80\% (pink). Furthermore, we plot 17
simulated elliptical galaxies formed from multiple mergers, in the
framework of a cosmological simulation: e) the most massive merger
remnants, re-simulated with twice the spatial resolution of the
original DM box (light-green), f) the less massive remnants,
re-simulated with four times the original resolution of DM
particles (dark-green), and g) four companion ellipticals, which
are substructures within larger haloes (yellow). R13 referred to
simulations (e--f) as CosmoZoom Ellipticals,  and simulations (g)
as CosmoZoom Companions. All simulations have been performed by
R13 with modified versions of the parallel TreePM-SPH-code
GADGET-2, including the effect of star formation and feedback from
SNe's, assuming a Salpeter IMF. Black hole (BH) growth and
feedback are included in the binary merger simulations only, while
CosmoZoom simulations do not include any BH treatment.

Several simulated galaxies (in particular the most massive
CosmoZoom Ellipticals) have masses larger than the range covered
by our data, hampering a direct comparison to our results.
Therefore, we focus the comparison on objects having $\log \mst
\lsim 11.8$ in Fig.~\ref{fig:simulations}. All binary mergers in
this mass range are in good agreement with our results in both the
$\alpha_{av}$--\mst\ and $\alpha_{av}$--\fdm\ plots. The same
result holds when comparing simulations to the trends for ETGs
with a Salpeter-like IMF normalization (see
Sec.~\ref{sec:literature}), i.e. the same IMF as in R13. On the
contrary, CosmoZoom galaxies, in the mass range from $\log\mst
\sim11.3$ to $\log\mst\sim11.8$ (green and yellow dots in the
Figure) have systematically steeper slopes, at a given stellar
mass, than our data, which is more consistent with slopes for a
constant \ML\ profile (i.e. the cyan region in the Figure). A
similar discrepancy exists with respect to \fdm , although in the
$\alpha_{av}$--\fdm\ plane, the deviation of CosmoZoom galaxies
from our fiducial trends (black curves) is smaller than in the
$\alpha_{av}$--\mst\ diagram. This is due to the fact that, at
fixed stellar mass, CosmoZoom galaxies also have lower DM
fractions than real galaxies.

We argue that the excellent agreement found between our results
and the predictions of binary merger simulations can be due to the
inclusion of BH feedback in them, which is more efficient than
SN-feedback in suppressing star formation
(\citealt{Tortora+09AGN}; \citealt{Martizzi+14}), producing less
stellar mass in the galaxy centre, and nearly isothermal total
mass profiles, in agreement with the observed trends.

Although the CosmoZoom simulations are offset with respect to the
observed trends in the $\alpha_{av}$--\mst\ diagram, they exhibit
a similar trend as in the data, with mass density slope increasing
(becoming shallower) with galaxy mass. The existence of such trend
can be explained by a smaller amount of dissipation during the
formation of high-, relative to low-, mass galaxies. During a
merger, gas dissipates its kinetic energy, falling into the galaxy
center and forming new stars. Therefore, a higher level of
dissipation leads to a more prominent contribution from newly
formed  stars to the total mass density in the center, steepening
the total density slope, as observed in low- relative to high-mass
(both observed and simulated) ETGs. The existence of a strong
correlation between density slope and radius
(Section~\ref{sec:results}) also supports this interpretation, as
dissipation would favour the formation of a smaller size system.

\section{Summary and Conclusions}\label{sec:conclusions}

In the present work, we have analyzed the stellar and DM
distribution in the central regions of ETGs, using a large sample
of nearby galaxies from the SPIDER sample (\citealt{SPIDER-I}), as
well as a complementary dataset of \atlas3d\ ETGs
(\citealt{Cappellari+11_ATLAS3D_I}). We have compared our findings
to independent results from the literature, and predictions from
numerical simulations. We have modeled each galaxy with two
components, a S\'ersic (de Vaucouleurs) profile for the SPIDER
(\atlas3d) sample plus a variety of viable profiles for the DM
distribution.  Our reference model is based on an NFW (DM) plus a
S\'ersic (stars) component, assuming a concentration--virial mass
relation from simulations~\citep{Maccio+08} and the virial to
stellar mass relation of~\citet{Moster+10}. Assuming circular
symmetry, no rotation, and neglecting radial gradients of the
stellar mass-to-light ratio, \Yst , in ETGs, we derive the only
free parameter of the model, the \Yst , from the central velocity
dispersion, \sigAp\ and \sige, of each galaxy. None of these
assumptions is found to affect significantly our results. From the
two-component models, we derive the total mass density slope in
the central regions of ETGs, and analyze its correlation with
several galaxy parameters, i.e. \sige, \mst, \Re , and $n$. Our
analysis (i) extends, with an independent approach, down to $\mst
\sim 10^{10} \rm \Msun$, the results of gravitational lensing
studies of massive galaxies (\citealt{Bolton+06_SLACSI};
\citealt{Bolton+08_SLACSV}; \citealt{Auger+09_SLACSIX};
\citealt{Auger+10_SLACSX}); (ii) complements previous work (e.g.
\citealt{Humphrey_Buote10}; \citealt{Dutton_Treu14};
\citealt{Chae+14}; \citealt{Oguri+14}) by targeting two
independent, large, samples of ETGs, and using a better tracer
(the K-band light) of the stellar mass distribution in galaxies;
and (iii) investigates the impact of a variety of modeling
ingredients on the inferred \Yst\ and central mass density slopes.

Our results can be summarized as follows:
\begin{description}
\item[--] Consistent with our previous work~\citep{TRN13_SPIDER_IMF}, we find that
ETGs at high \sige\ have larger \Yst\ than that expected for a
Chabrier IMF when fitting either colours or galaxy spectra with
stellar population models, i.e. that the mismatch parameter,
\dimf$=$\Yst$/$\Yst$^{Chab}$, becomes significantly larger than
one at high \sige. This result can be interpreted as a systematic
variation of the IMF normalization (i.e. the amount of stellar
mass in the IMF) with \sige . In the present work, we find that
\dimf\ also increases with stellar mass and \Re\ (but to less
extent than the trend with \sige), while it decreases with the
S\'ersic $n$. Using \atlas3d\ sample we find, on average, larger
\Yst\ and shallower correlations with all the parameters. These
results are consistent with studies of gravity-sensitive features
in ETGs, finding that the IMF becomes bottom-heavier than a
``standard'', MW-like, distribution in high-, relative to
low-\sige\ ETGs (e.g.~\citealt{Ferreras+13,
LaBarbera+13_SPIDERVIII_IMF, Spiniello+14}). At low \sige\ ($\sim
100$~$\rm km \, s^{-1}$), the value of \dimf\ ($\sim 1$) implies a
MW-like IMF normalization, consistent with results for late-type
galaxies (\citealt{Sonnenfeld+12}; \citealt{Brewer+12}), and the
combined lensing and stellar population analysis of \cite{FSB08,
Ferreras+10}. The trends of \dimf\ hold for all DM profiles
explored in this work, with lower \dimf's for contracted halo and
``high-concentration'' models (the latter being based on the
\cvir\ -- \Mvir\ relation from LFS12).

\item[--] For our reference model (see above), the total mass density slope in the centre of ETGs increases (becoming less negative)
with galaxy mass and galaxy size. For the \atlas3d\ sample we find
consistent results to those for SPIDER ETGs, although the trend
with mass is steeper for the latter. In more detail, we find that
low-mass (small) ETGs have slopes consistent with those for
constant-\ML\ profiles, while massive (large \Re ) systems have a
nearly isothermal density slope ($=-2$), consistent with
gravitational lensing results (e.g., \citealt{Gavazzi+07_SLACSIV};
\citealt{Auger+10_SLACSX}). The trends of  mass density slope are
consistent with independent results from the literature
(\citealt{Humphrey_Buote10}; \citealt{Dutton_Treu14}). In terms of
central velocity dispersion, the density slope decreases with
central velocity dispersion (but to less extent than the amount of
variation with \Re ), while it exhibits a double-value behaviour
with the S\'ersic $n$, increasing at both low and high $n$, with a
minimum for $n$ between $3$ and $5$ (depending on the method to
compute the slope).

\item[--] The trends of mass density slope are the same for NFW and contracted-NFW models,
and do not change when assuming a fixed virial mass (and
concentration) for all galaxies (rather than a virial to stellar
mass relation, such as that of~\citealt{Maccio+08}). When adopting
a Burkert profile, the slope tends to be more constant as a
function of all galaxy parameters explored. However, for the most
massive ETGs, the ``light'' haloes described by Burkert models
seem to be rejected by lensing results
(\citealt{Auger+09_SLACSIX}; see also \citealt{CT10}).

\item[--] Using a \cvir\ -- \Mvir\ relation from observations (LFS12) rather than simulations ~\citep{Maccio+08}
affects significantly some trends of density slope  with galaxy paramaters. In particular, while the slope
keeps increasing with galaxy radius  also for ``high-concentration'' models (with \cvir\ -- \Mvir\ from LFS12),
the trends with mass become flatter in this case. On the other hand, the trends with central velocity dispersion
are the same for all models.

\end{description}

Our results corroborate a picture whereby the total mass density
profile in the central regions of ETGs is ``non-homologous'',
approaching a constant-\ML\ distribution at low mass -- where
stars dominate the total mass budget in the center --, and an
isothermal profile in the most massive ETGs, whose central regions
are more DM dominated. The fact that the mass density slope of
groups and clusters of galaxies seems to be shallower than that of
massive galaxies (e.g. \citealt{Sand+08};
\citealt{Humphrey_Buote10}) further supports the ``non-homology''
of the total mass distribution of galactic systems.

To understand the implications of our findings in the framework of
galaxy assembly, we have also compared our results, i.e. the
trends of total mass density slope, with simulation predictions
from~R13. The comparison indicates that BH growth and feedback are
essential ingredients during the formation of ETGs, as only
simulations including them are able to reproduce the mass density
slope, DM fraction, and stellar mass we have measured in the
central regions of ETGs. Also, we find that both observations and
simulations predict an increase of the total mass density slope
with galaxy mass. We argue that this trend is because gas
dissipation has been more important during the formation of low-,
relative to high-, mass galaxies. In such a picture, a steep
profile is due to the formation of new stars inwards, as the gas,
dissipating  its kinetic energy, falls into the galaxy central
regions, while gas-poor mergers tend to make the slopes
isothermal.

The present work shows that observations and simulations are now
converging to provide a consistent characterization of the
luminous and DM components in the central regions of ETGs.
Nevertheless, important questions remain still open, like the
discrepancy between halo concentration and mass from N-body
simulations, and those obtained from lensing studies. In the
future, it will be important to extend our results to the
outermost regions of these galaxies, taking advantage of data
covering a wide galactocentric baseline (e.g. kinematical tracers
as planetary nebulae or globular clusters,
\citealt{Romanowsky+09}; \citealt{Napolitano+09_PNS};
\citealt{Napolitano+11_PNS}; \citealt{Pota+13_SLUGGS}) and
accounting for radial gradients of the stellar IMF
(\citealt{Martin-Navarro+14}; \citealt{Pastorello+14}). From the
theoretical viewpoint, it will be also interesting to explore
phenomenologically-motivated models for the mass distribution in
galaxies (e.g. \citealt{Zhao97}; \citealt{Tortora+07};
\citealt{Cardone+09}), as well as alternative theories with
modified gravity,  like MOND (\citealt{Milgrom83, Milgrom83b};
\citealt{Cardone+11MOND}; \citealt{Tortora+14_MOND}) and
modifications of the Einstein theory (e.g. $f(R)$,
\citealt{Lubini+11}; \citealt{Napolitano+12_fR}).


\section*{Acknowledgments}

We thank the referee for the kind report. CT has received funding
from the European Union Seventh Framework Programme
(FP7/2007-2013) under grant agreement n. 267251.


\bibliographystyle{mn2e}   


\end{document}